\documentclass[showpacs,12pt]{iopart}
\usepackage[pdftex]{graphicx}
\usepackage{hyperref}
\pdfoutput=1
\hypersetup{pdfauthor={some author},pdftitle={eye-catching title}}
\ifpdf

\begin{document}

\title{Configurational instability at the excited impurity ions in alkaline earth fluorites}
\author{A.V. Egranov$^{1,2}$ and T.Yu Sizova$^1$ }

\address{$^1$ Vinogradov Institute of Geochemistry, Russian Academy of Sciences, Favorskii street 1a, 664033 Irkutsk, Russia}
\address{$^2$ Irkutsk State University, Faculty of Physics,  Gagarin Blvd. 20, 664003 Irkutsk, Russia}
 \ead{alegra@igc.irk.ru}

\begin{abstract}
The formation of intrinsic defects by ionizing radiation in some ionic crystals in the anion sublattice is only associated with the instability of the self-trapped exciton up to now. In this paper we propose a new mechanism for the formation of the defects in the anion sublattice associated with the Janh-Teller instability occurring near the cation impurities in the excited state. The instability occurs when the degenerate excited state of the impurity ion is localized in the conduction band. We believe that the configuration interaction between the discrete impurity level and host continuum (effect Fano) plays an important role in this process. 

\end{abstract}

\maketitle

\section{Introduction}
Frenkel pairs comprising a vacancy and an interstitial in the anion sublattice are generated by ionizing radiation in some insulators in which electron-lattice coupling is strong. The mechanism of generation of Frenkel pairs in several solids, typically in alkali halides has been studied extensively and it has been generally accepted that they are generated as a result of the adiabatic instability of self-trapped excitons \cite{Toyozawa:2003, Lushchik:1989, Song:1993, Itoh:2001, Itoh:1990}. These processes have been studied well enough in alkali halide crystals. These crystals are easily colored by  ionizing radiation and defects responsible for this coloration were called color centers.

It has been found that alkaline earth fluoride crystals which have not been deliberately doped with impurities are much less susceptible to coloration at room temperature by x irradiation than most alkali halides. Several review articles and books have dealt with this subject,  but the situation is still not clearly understood and no direct analogy appears to exist with the more widely studied alkali halides \cite{Song:1993, Itoh:2001, Hayes:1974}. The behavior of x-irradiated alkaline earth fluorides doped with some cation impurities differs markedly from that of the pure crystals. The crystals can be easily colored by ionizing radiation.  The defects formed are not only associated with changes in the valence state of the impurities, but also with the intrinsic damages in the anion sublattice \cite{Egranov:2008, Egranov:2008a}.

The photochromic centers are produced either by x-irradiation or by additive coloration (by heating the
crystals in a calcium atmosphere) of CaF$_2$ crystals doped with certain rare earths ions (La, Ce, Gd, Tb and Lu) or yttrium. On the basis of optical and electron paramagnetic resonance (EPR) work, as well as theoretical investigations, it has been suggested that the ionized and thermally stable photochromic centers in CaF$_2$  crystals consist of one and two electrons bound at an anion vacancy adjacent to a trivalent impurity cation \cite{Anderson:1971, Staebler:1971, Alig:1971} and they were called as PC$^+$ (Fig. ~\ref{formation}) and PC respectively. It has been found that photochromic centers can be formed by the shallower trivalent traps, e.g. by Y, La, Ce, Gd, Tb and Lu where low third ionization potentials are expected; the deeper traps with the higher ionization potentials will form divalent ions. 

In the review by Hayes and  Staebler at 1974 \cite{Hayes:1974}, it was concluded that "the formation mechanism of the PC centre is not yet understood, and little known of the mechanism by which the centre is optically ionized. There is clearly a need for further exploration of these interesting systems." So far, little has changed. In this paper we want to show that the formation of the anionic intrinsic defects (the photochromic centres) is due to the instability at the excited cationic impurities.

\section{Photochromic centers}
Two experimental results obtained recently could help us to understand the possesses which occurs under radiation excitation in the crystals \cite{Bugaenko:2008}

- x-irradiation at 77 K of the CaF$_2$ and SrF$_2$ crystals doped with the impurities which can form the photochromic centers, results in creation PC$^+$ and V$_k$ centers (self-trapped hole which have the structure of molecular ion - F$_2$$^-$).

- efficiency of the formation of the photochromic centers decreases from CaF$_2$ to SrF$_2$ and in BaF$_2$ crystals the centers are not created under x-irradiation.

Anion vacancy ({$\alpha$}-centre) which is part of the PC$^+$ center is an intrinsic radiation defect usually arising from radiationless decay of self-trapped excitons. Creation of anion vacancy at nearest-neighbor of trivalent rare earth ions at 77 K is unexpected process and as far as we know it is the unique case. The defects having an anion vacancy at the nearest-neighbor position are created at the temperatures higher than onset of the motion of anion vacancies. Thus F$_A$ centers are fully converted to (F$_2$$^+$)$_A$ centers in CaF$_2$-Na crystals according to thermal conversion process F$_A$ + {$\alpha$} ${\rightarrow}$  (F$_2$$^+$)$_A$ \cite{Tijero:1990} or in SrF$_2$ and CaF$_2$ crystal doped with Cd the reduction of symmetry of the Cd$^+$ center is due to the attachment of anion vacancy in the nearest neighbor site \cite{Egranov:2008, Egranov:2008a}. Onset of the motion of an anion vacancy in CaF$_2$ crystal is at about 200 K \cite{Tijero:1990}. 

\begin{figure}
\centering
\includegraphics[width=5.2in, height=3.6in]{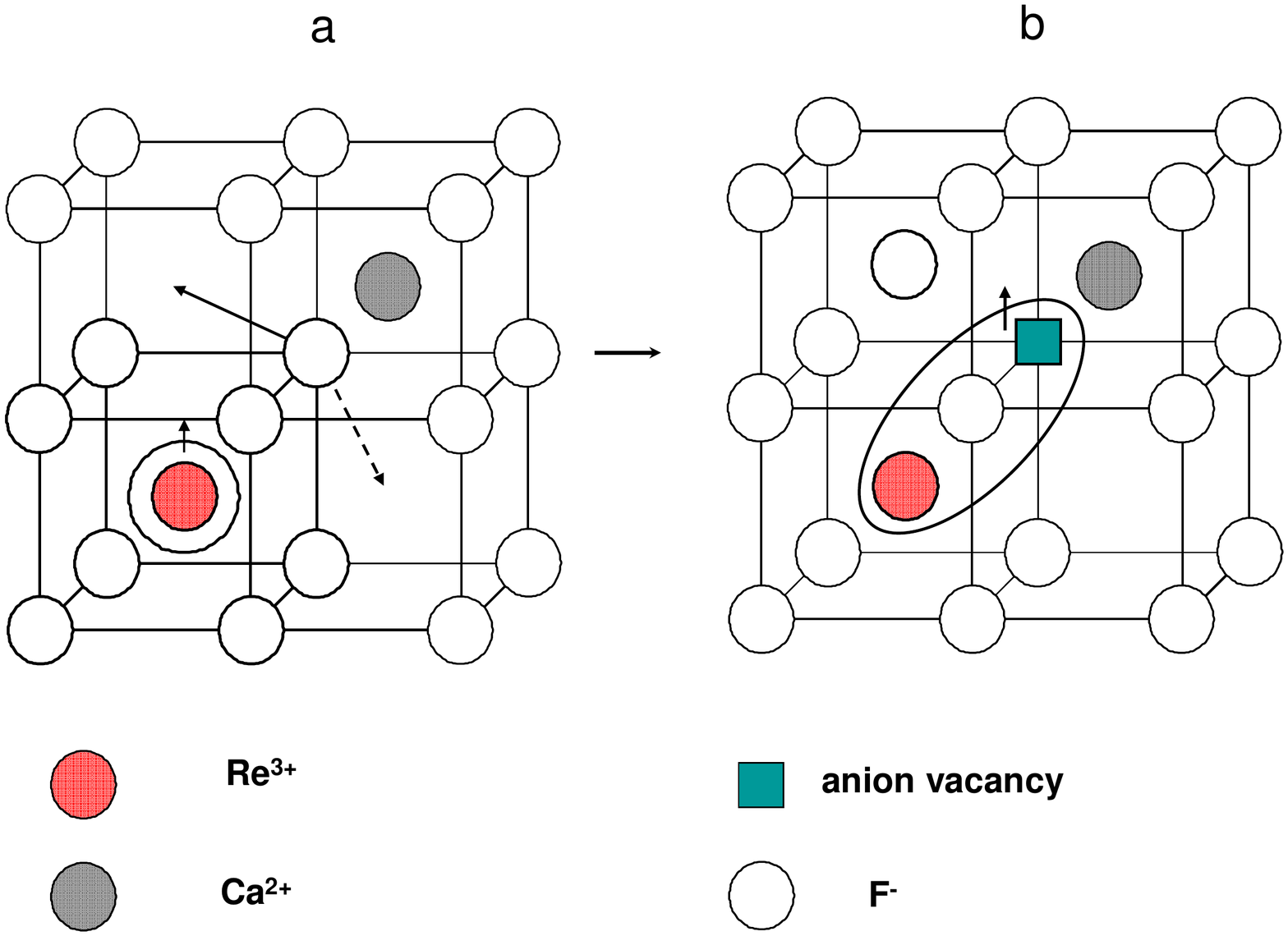}
\caption{Formation of the PC$^+$ centres. (a) Electron capture by the trivalent rare earth ion having O$_h$ symmetry leads to the formation of the divalent rare earth ion in an excited degenerate state. This state is unstable to a Jahn-Teller distortion. The distortion leading to PC$^+$ formation involves a replacement of one of the nearest-neighbor F$^-$ ion into an intestitial position and the relaxation of the electron from the excited state into  the anion vacancy. (b) The result is a PC$^+$ center perturbed by interstitial fluorine ion, which prevents the subsequent capture of an additional electron by this complex.}
\label{formation}
\end{figure}

This fact indicates that the instability which leading to the formation of anion vacancy at nearest-neighbor of trivalent rare earth ion is not associated with the decay of self-trapped excitons in a regular lattice because they are immobile at this temperature. Ionizing radiation produces in a crystal free electrons and holes. The holes are self-trapped at 77 K with the formation of a quasimolecule F$_2$$^-$ and the electrons can be trapped by the impurity ions. It is reasonable to assume that the instability occurs when trivalent rare earth ion captures electrons to form a divalent ion in an excited state. Catlow \cite{Catlow:1979} first suggested the possibility of such instability at the excited cation ion after an electron capture.

Most of the previous studies of the photochromic centers were carried out on additively colored crystals. However, there are a number of works on coloration of these crystals by x-irradiation at low temperatures \cite{Scouler:1960, Merz:1967, O'Connor:1964}. It was noted \cite{Hayes:1974, Scouler:1960} that the absorption caused by ultraviolet excitation of PC centres at 78 K in additively colored CaF$_2$-Y is similar to that produced by x-irradiation of CaF$_2$-Y at 78 K. The results show unequivocally that x- irradiation at 78 creates PC$^+$ centers in CaF$_2$-Y, but this conclusion is why it has not been made. The situation is very much complicated by the strong overlapping of the PC$^+$ centers and the divalent ions absorption.
The optical absorption spectra of divalent rare earths in CaF$_2$ after x-irradiation at 77 K have been given by Merz and Pershan \cite{Merz:1967}. Only a small part of trivalent ion converted to divalent by x-ray irradiation. Therefore, the divalent ions have a relatively weak absorption. However, the absorption spectra for  La, Ce, Gd and Tb (Lu has not been studied) is somewhat different from the rest. The "divalent bands" at wavelengths longer than 400 nm for La, Ce, Gd, and Tb are much stronger than the absorption bands in this region of the spectrum for the other ions. The existence of these anomalous bands, as well as the hole absorption at 310 nm (V$_k$ - centres), casts some doubt on the blanket assertion that the radiation-induced absorption curves are due solely to the divalent rare earths. 
Comparison between these results and those obtained later \cite{Staebler:1971} indicates that these absorption spectra are associated with PC$^+$ centers.

From the above considerations it is clear that, despite the experimental results indicating the formation of PC$^+$ centers at low temperatures by x-irradiation, it was assumed that the formation of the PC centers arises from thermally activated motion of anion vacancies \cite{Hayes:1974}.

Since the instability, leading to the formation of anion vacancies, is associated with the distortion of the nearest anion environment at the excited impurity divalent ion, it is reasonable to assume that this distortion is conditioned by the Jahn-Teller effect. The Re$^2$$^+$ impurity occupies a substitutional site of metal in the fluorite-type structure crystal. The site symmetry is cubic (point group 0$_h$) with eightfold coordination. At 77 K Re$^2$$^+$ ion producing by electron capture of Re$^3$$^+$ after x-irradiation may be perturbed by interstitial F$^-$ ion locating in nearest-neighbour (nn) positions, thus leaving this dipolar (RE$^3$$^+$-F$^-$) complex with tetragonal (C$_4$$_v$) symmetry. There is evidence that, at least at low temperatures, only those lanthanide ions with non-local compensator (cubic sites ) can be reduced to the divalent state. Thus the spectra of the thermoluminescence corresponds to the fluorescence of the trivalent ion in sites of cubic symmetry for the low-temperature glow peaks, and tetragonal symmetry above room temperature \cite{Merz:1967}.

For free ions the ground configuration of La$^{2+}$, Gd$^{2+}$ and Y$^{2+}$ is 5d$^1$, for Ce$^2$$^+$, Tb$^2$$^+$, Lu$^2$$^+$ is f$^n$ but f$^n$$^-$$^1$5d and f$^n$ states are close to each other. For crystals with the fluorite-type structure, the crystal field acting at the Re$^2$$^+$ ion possesses cubic symmetry with eightfold coordination which splits the fivefold orbital degeneracy of the d-energy level into doubly-degenerate ($^2$E$_g$) and threefold degenerate ($^2$T$_2$$_g$) energy levels with the $^2$E$_g$ level lowest. The spin-orbit coupling splits the $^2$T$_2$$_g$ energy level into the two states  \cite{Manthey:1973} and particular quenches the Jahn-Teller distortion. In cubic symmetry the spin-orbit coupling does not split the $^2$E$_g$ state.

\begin{figure}
\centering
\includegraphics[width=5.2in, height=3.6in]{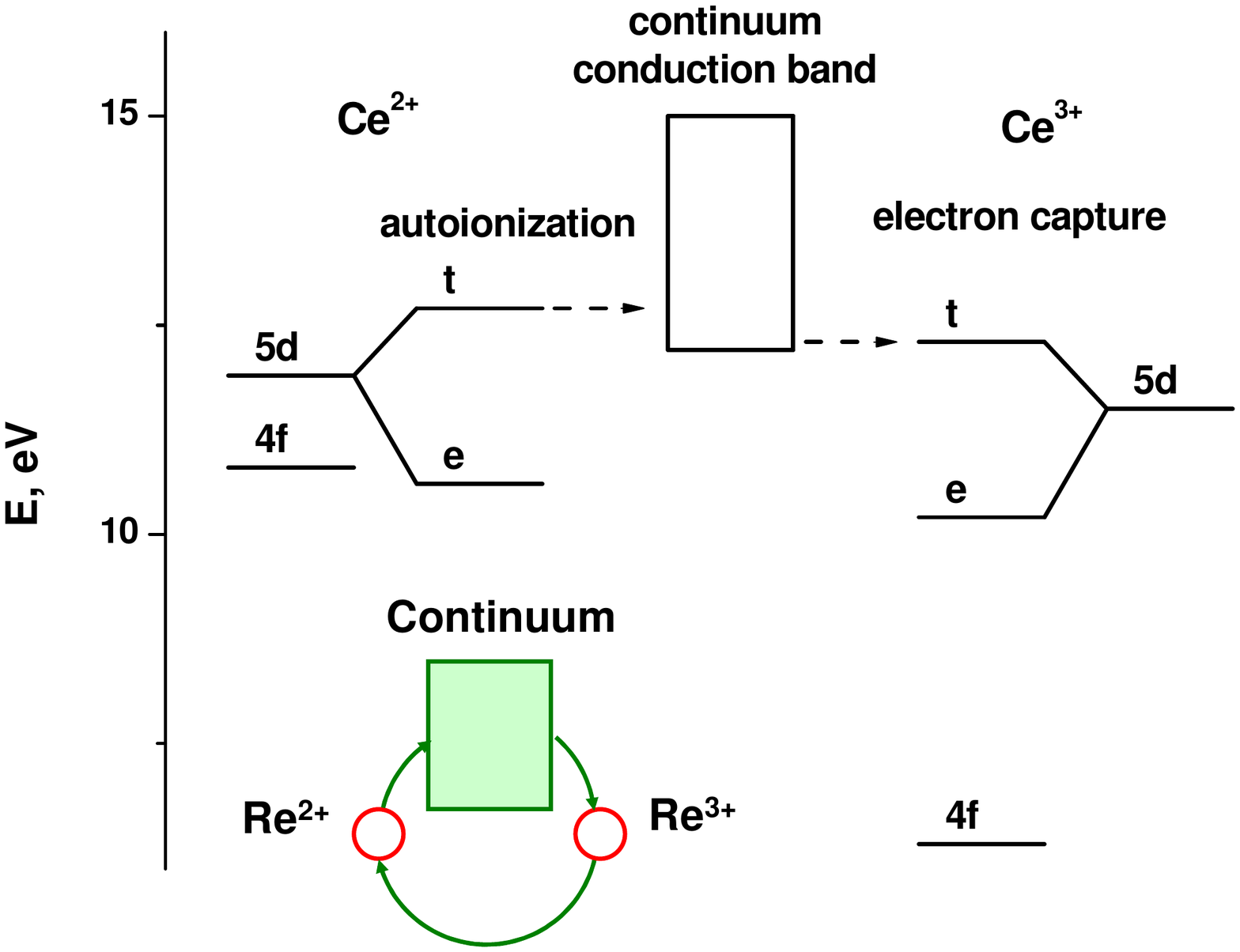}
\caption{Energy levels of Ce$^2$$^+$ and Ce$^3$$^+$ in cubic F$^-$ coordination (schematic) relative to the conduction band in CaF$_2$. Spin-orbit coupling is neglected. The inset shows the cycle associated with the valence fluctuation of the rare earth ion.}
\label{state}
\end{figure}

Experimental and theoretical evaluations of the location of the energy levels of rare earth ions in CaF$_2$, SrF$_2$, and BaF$_2$ relative to the conduction band states of the host crystal show that threefold degenerate ($^2$T$_2$$_g$) energy levels of the divalent rare earth ions are in conduction band of the crystals \cite{Pedrini:1986, Dorenbos:2003}. For example, the energy levels of Ce$^2$$^+$ and Ce$^3$$^+$ in cubic F$^-$ coordination relative to the conduction band in CaF$_2$ is shown in Fig.~\ref{state}.  Since the $^2$T$_{2g}$ orbital of the 5d electron is orbitally degenerate, it will interact with e - or t - symmetry vibrational modes of the eight neighboring fluorine ions and this interaction leads to the Jahn-Teller effect in its many forms. Jahn-Teller distortion arises when trivalent rare earth capture electron forming divalent rare earth ion in excited $^2$T$_2$$_g$ state. The distortion leading to PC$^+$ formation involves a motion of the nearest-neighbor F- ion into an interstitial position and the relaxation of the electron from the excited state into  the anion vacancy. The PC$^+$ center is disturbed by interstitial fluoride ion (Fig. ~\ref{formation}), which prevents the subsequent capture of an additional electron by this complex. At about room temperature the interstitial fluoride ion can be removed from the neighborhood of the PC$^+$ center and capture of an additional electron by the PC$^+$ center can occur.  As a result, x-irradiation at room temperature leads to the formation of PC center having two electrons. 

The PC$^+$ center is lower-energy equilibrium complex than the divalent state for the rare earth ions with low third ionization potencials. Consequently, the formation of an anion vacancy lowers the energy of the system and symmetry reduces from O$_h$ to C$_3$$_v$. It should be noted that none of these rare earth metals exhibits a stable +2 state in any of its compounds and consideration of factors such as their ionization energies suggest they are unlikely to form stable compounds in this state \cite{Cotton:2006}. 

For some divalent rare earth ions having the d$^1$($^2$E$_g$) configuration (La$^2$$^+$, Gd$^2$$^+$ and Y$^2$$^+$) in the ground state the Jahn-Teller effect have been observed by EPR \cite{Herrington:1970, Herrington:1971, Herrington:1973,  Bill:1984, Bill:1986, Bill:1989}, but this distortion does not leads to the formation anion vacancy. Essential requirement for the Jahn-Teller distortion leading to the PC$^+$ formation is to localization of $^2$T$_2$$_g$ state of the divalent rare earth ion in conduction band. The excited state $^2$T$_2$$_g$ of Re$^2$$^+$ embedded in the continuum energy state which can autoionize can also decay to a lower-lying state forming the divalent ion in the ground state or can distort by Janh-Teller effect with formation of the PC$^+$ center. The formation of the divalent ions has a low efficiency, regardless of the efficiency of formation of the photochromic centers in different crystals. The independence of the formation of the divalent ions and the photochromic centers provided via autoionization. PC and PC$^+$ centers can be efficiently created in CaF$_2$ by x-irradiation, can be inefficiently created in SrF$_2$ and cannot be created in BaF$_2$ crystals \cite{Bugaenko:2008}. This fact may be explained by decreasing of the covalent interaction between Re$^{2+}$ and eight F$^-$ neighbors with increasing of the lattice  parameter from CaF$_2$ to BaF$_2$. 

\section{"Anomalous" emission} 
Let us consider another case that as we believe is close to the former. For the divalent lanthanides (Re$^2$$^+$) in alkaline earth fluorides, under photoexcitation d${\rightarrow}$ f emission has been only observed for Eu$^2$$^+$ and Yb$^2$$^+$ ions. In the other divalent lanthanides, again relaxation to lower-lying 4f$^n$ levels quenches the d${\rightarrow}$ f emission \cite{Dorenbos:2003a}. For Yb$^2$$^+$ in CaF$_2$, SrF$_2$, BaF$_2$, Eu$^2$$^+$ in BaF$_2$ has been observed the broadband emissions with large Stokes shift which is due to strong distortion of the environment and these emissions were called as "anomalous" emission. 
At first the anomaly of this emission was explained as arising from the Jahn-Teller distortion \cite{Kaplyanskii:1976}. It was shown subsequently that the emission is observed when doubly-degenerate d($^2$E$_g$) level from which electron delay to ground state via emitted photon must be localized in conduction band \cite{McClure:1985, Pedrini:2007, Dorenbos:2003a}. It is now accepted that the location of the 5d levels relative to conduction band state and the presence of the "anomalous" emission are related to each other. 

It has been suggested that the "anomalous" luminescence results from the radiative decay of the so called  impurity trapped exciton \cite{McClure:1985, Pedrini:2007, Dorenbos:2003a}. The structure of the impurity trapped exciton is of the hole localized on the luminescent ion and the electron delocalized over the surrounding cations. The impurity trapped exciton in SrF$_2$ :Yb$^2$$^+$, for example, consists of a Yb$^3$$^+$ core with the electron delocalized over the 12 nearestneighbors Sr$^2$$^+$ ions \cite{McClure:1985}. 

The large Stokes shift observed has been explained on the basis of the distortion induced in the excited state with respect to the ground state geometry of the impurity trapped exciton. This effect is a consequence of the different radii between Re$^3$$^+$ and Re$^2$$^+$. After creation of the impurity trapped exciton, F$^-$ ligands collapse to wards Re$^3$$^+$. The model of the impurity trapped exciton  is based on the important, however, only single experimental result associated with localization of the excited state in conduction band and ignores the facts received early.

In this model the distortion has the cubic symmetry. However, the data obtained by Kaplyanskii \cite{Kaplyanskii:1976} from polarized luminescence and uniaxial stress studies have been shown that the distortion has the symmetry lower than cubic. From the polarization measurements of the luminescence the authors \cite{Moine:1990} of the model of the impurity trapped exciton concluded also that the luminescence center has C$_4$ orientation.

The "anomalous" emission due to the ionization process is not confined to the Eu$^2$$^+$ and Yb$^2$$^+$ ions and the d${\rightarrow}$f transitions (let via autoionization) but has been observed in the luminescence of ions such as Tm$^2$$^+$ \cite{McClure:1985}, Ce$^3$$^+$ \cite{Birowosuto:2006}, Bi$^3$$^+$ \cite{Srivastava:1999} (p${\rightarrow}$s transition) and Cu$^+$ \cite{Payne:1984} ($^3$E$_g$(d$^9$s)${\rightarrow}$ $^1$A$_1$$_g$(d$^1$$^0$) transition). 

However, unlike the BaF$_2$:Eu$^2$$^+$ system, the luminescence of BaF$_2$:Sm$^2$$^+$ appears to be "normal"; it is dominated by the sharp f$^6$($^5$D$_0$) ${\rightarrow}$ f$^6$($^7$F$_1$) transition as for the luminescence in SrF$_2$: Sm$^2$$^+$, in spite of the f$^6$($^5$D$_0$) state is located in the conduction band \cite{Fuller:1990}. The $^5$D$_0$ state in BaF$_2$:Sm$^2$$^+$ is just 1000 cm$^-$$^1$ below the lowest f$^5$d(e$_g$) state.

The shielding character of the 4f orbitals by the filled 5s$^2$ and 5p$^6$ orbitals determines the weak interaction with the neighbouring ions. This explains the small influence of the host lattice and the sharp lines observed in spectra based on f - f transitions and the quench the Jahn-Teller effect if electron occupies the f - orbital.
We thought that the processes of the formation of the photochromic centers and the "anomalous" luminescence are similar each other. In both cases it is needed the localization of excited d state in conduction band leading to the distortion of the environment. The distinction is that  the formation of PC$^+$ centers is energetically favorable for the shallow traps (Y$^2$$^+$, La$^2$$^+$, Ce$^2$$^+$, Gd$^2$$^+$, Tb$^2$$^+$, Lu$^2$$+$),  whereas for the deep traps (Eu$^2$$^+$, Yb$^2$$^+$) the divalent state is energetically more favorable.

\section{Conclusion}
The fact that the "anomalous" emission and the formation of the photochromic centers occur from the d- degenerate states of the divalent rare earth ion embedded in the continuum energy state indicate that the configuration interaction between the discrete level and continuum plays an important role in these processes. Fano investigated the stationary states with configuration mixing under conditions of autoionization and he pointed out the basic physics of the quantum interference phenomenon between a discrete level and a continuum \cite{Fano:1961}. The lifetimes of these excited states due to allowed autoionization (maybe via phonon - vibrational autoionization ) (typically 10$^-$$^1$$^5$-10$^-$$^1$$^3$ sec) are shorter by many orders of magnitude than the lifetimes due to allowed emission transitions (typically 10$^-$$^9$ sec).  In the case of BaF$_2$:Eu$^2$$^+$, the spectrum of the group of f$^6$d excited states from 3 to 4 eV is broadened by some radiationless process suggested to be autoionization. The lifetime was found to be 25 fs for the f$^6$d excited states and 70 fs for the f$^7$($^6$P$_7/2$) excited state which has comparable energy and is observed via two photon spectroscopy \cite{Fuller:1987}. After the autoionization the electron can be then recaptured due to coulomb interaction with the excess charge of the trivalent rare earth ion. This means that the valence of the rare earth ion (Re) wills fluctuate between the states Re$^2$$^+$ and Re$^3$$^+$. It is possible that this valence fluctuation creates the resonant electron-phonon interaction that leads to the enhancement the Jahn-Teller distortion. However there is the need of a theoretical exploration of this effect (maybe by using Barry phase cycle \cite{Berry:1984}, Fig. ~\ref{state}). The Fano type interference can occur between two cannels of the autoionizations either by the direct hopping or via phonon.

It is noteworthy to mention, that in the rare-earth compound such as SmB$_6$ which belongs to the class of intermediate-valence systems maybe the similar effect has been found. Jahn-Teller effect on Er$^3$$^+$, Sm$^3$$^+$, Gd$^3$$^+$  ions was observed only in a compound with fluctuating valence SmB$_6$ but neither in an isomorphic compound LaB$_6$ where La has a constant valence nor in the semiconductors BaB$_6$, YbB$_6$, CaB$_6$, etc. \cite{Sturm:1985, Al'tshuler:2003}.  Authors concluded that strong electron-phonon interaction which is essential for observation of the Jahn-Teller effect plays a significant role in the nature of fluctuating valence.

\section*{Acknowledgments}
This work was partially supported by grant 11-02-00717-a from Russian Foundation for Basic Research (RFBR) and  by the Lavrentjev's  grant of the Siberian Branch of Russian Academy of Science 7.11.

\section*{References}
\bibliographystyle{utphys}
\bibliography{Egranov1}

\end{document}